\documentclass[%
 aip,
 amsmath,amssymb,
 reprint,%
]{revtex4-1}

\usepackage{graphicx}
\usepackage{dcolumn}
\usepackage{bm}
\usepackage{color}
\usepackage{ulem}
\usepackage{threeparttable}
\usepackage{longtable}
\usepackage{afterpage}
\usepackage{multirow}
\usepackage{siunitx}
\usepackage{comment}
\usepackage{amsmath}
\usepackage[T1]{fontenc}
\usepackage[utf8]{inputenc}
\usepackage{mathptmx}
\usepackage{etoolbox}

\makeatletter
\def\@email#1#2{%
 \endgroup
 \patchcmd{\titleblock@produce}
  {\frontmatter@RRAPformat}
  {\frontmatter@RRAPformat{\produce@RRAP{*#1\href{mailto:#2}{#2}}}\frontmatter@RRAPformat}
  {}{}
}%
\makeatother

\begin{document}

\title{Interfacial Mixing Effect in a Promising Skyrmionic Material: Ferrimagnetic Mn$_4$N}
\author{Chung T. Ma}
    \email{ctm7sf@virginia.edu}
\affiliation{Department of Physics, University of Virginia, Charlottesville, Virginia, 22904, USA}
\author{Wei Zhou}
\affiliation{Department of Physics, University of Virginia, Charlottesville, Virginia, 22904, USA}
\author{Brian J. Kirby}
\affiliation{NIST Center for Neutron Research, National Institute of Standards and Technology, Gaithersburg, Maryland 20899, USA}
\author{S. Joseph Poon}
\affiliation{Department of Physics, University of Virginia, Charlottesville, Virginia, 22904, USA}
\affiliation{Department of Materials Science and Engineering, University of Virginia, Charlottesville, Virginia, 22904, USA}
\date{\today}

\begin{abstract}
Interfacial mixing of elements is a well-known phenomenon found in thin film deposition. For thin-film magnetic heterostructures, interfacial compositional inhomogeneities can have drastic effects on the resulting functionalities.  As such, care must be taken to characterize the compositional and magnetic properties of thin films intended for device use.  Recently, ferrimagnetic Mn$_4$N thin films have drawn considerable interest due to exhibiting perpendicular magnetic anisotropy, high domain-wall mobility, and good thermal stability. In this study, we employed X-ray photoelectron spectroscopy (XPS) and polarized neutron reflectometry (PNR) measurements to investigate the interfaces of an epitaxially-grown MgO/Mn$_4$N/Pt trilayer deposited at 450 $^{\circ}$C. XPS revealed the thickness of elemental mixing regions of near 5 nm at both interfaces. Using PNR, we found that these interfaces exhibit essentially zero net magnetization at room temperature. Despite the high-temperature deposition at 450 $^{\circ}$C, the thickness of mixing regions is comparable to those observed in magnetic films deposited at room temperature.  Micromagnetic simulations show that this interfacial mixing should not deter the robust formation of small skyrmions, consistent with a recent experiment. The results obtained are encouraging in terms of the potential of integrating thermally stable Mn$_4$N into future spintronic devices.
\end{abstract}

\maketitle

\section{introduction}

The interface plays an important role in the properties of thin-film heterostructures. For magnetic materials, many well-studied and important phenomena arise from interfacial effects, such as interfacial magnetic anisotropy \cite{Engel1991,Nakajima1998,yang2011,Dieny2017}, RKKY coupling in a synthetic antiferromagnet through a metal spacer \cite{Leal1998,parkin2003,Duine2018,Legrand2020}, and the interfacial Dzyaloshinskii-Moriya interaction (I-DMI) \cite{DZYALOSHINSKY1958,Moriya1960,Hrabec2014,yang2015,Belmeguenai2005,Quessab2020}. Furthermore, it has been reported that in many of these heterostructures, interfacial mixing and dead layers exist at the interfaces of magnetic and non-magnetic layers. Such interfacial compositional heterogeneities have led to varying magnetic properties, such as thickness-dependent anisotropy and magnetization \cite{carcia1990,Wang2006,Jang2010,Li2012,Hebler2016,Ma2018,Laureti_2021}. These present challenges in fabricating and developing devices with magnetic thin films, such as magnetic tunnel junctions \cite{ZHU2006}. Therefore, investigating the effects due to interfacial mixing in heterostructures is a crucial step in designing new materials for spintronics applications. 

Mn$_4$N thin films have recently garnered ¬much attention in the spintronic community \cite{Ghosh2021}. Mn$_4$N forms an anti-perovskite crystal structure \cite{Takei1962_Mn4N}. It is a ferrimagnetic metal with a Curie temperature of 710 K \cite{Meinert_2016}. Experiments have shown that Mn$_4$N exhibits perpendicular magnetic anisotropy (PMA) \cite{Yasutomi2014Mn,Kabara2015Mn4N,FOLEY2017Mnn4,Hirose2020Mn4N,HIROSE2020-1,Isogami2020Mn4n,Zhou2021_Mn4N} that is desirable for use as free and fixed layers in magnetic devices. Furthermore, these properties are resilient to high temperatures, at least up to 450 $^{\circ}$C. More recently, tunable magnetic skyrmions have been reported to be present in MgO/Mn$_4$N/Pt$_x$Cu$_{1-x}$ thin films \cite{Ma2021} wherein Pt$_x$Cu$_{1-x}$ is designed to provide a variable I-DMI that controls the size of the skyrmion. As such, Mn$_4$N can potentially be used as a building block in skyrmion-based memory devices. Nonetheless, in view of elemental mixing that tends to occur at the interfaces of thin-film heterostructures, the question remains as to what extent high-temperature deposition Mn$_4$N \cite{Yasutomi2014Mn,Kabara2015Mn4N,FOLEY2017Mnn4,Hirose2020Mn4N,HIROSE2020-1,Isogami2020Mn4n,Zhou2021_Mn4N} would adversely affect the functionalities of the Mn$_4$N based heterostructures. For example, the interface is known to play a significant role in the effectiveness of spin currents and spin-orbit torque between magnetic and non-magnetic layers \cite{Toka2015,Amin208,Jin2019,Li2021}. In a recent study, Mn$_4$N has been reported to have a 2 nm mixing layer from transmission electron microscopy \cite{HIROSE2020-1}. Thus, further quantitative studies of the interfacial properties of Mn$_4$N based heterostructure films is pivotal in exploring the feasibility of using Mn$_4$N in future devices. 

In this study, we investigated the interfaces of epitaxially-grown MgO/Mn$_4$N (17 nm)/Pt films, which were deposit at the same time, using X-ray photoelectron spectroscopy (XPS) and polarized neutron reflectometry (PNR). From XPS, mixing layers were revealed at both MgO/Mn$_4$N and Mn$_4$N/Pt interfaces. At the MgO/Mn$_4$N interface, an oxidation layer of 5 nm thick MgMnNO mixing layer was revealed. At the Mn$_4$N/Pt interface, a 5 nm thick MnPtNO mixing layer was found. Analysis of the PNR data revealed that both top and bottom mixing layers exhibit essentially zero magnetization. The 5 nm thickness of the mixed layer at the MgO/Mn$_4$N interface is comparable to other room temperature depositions \cite{Hebler2016,Ma2018}, despite Mn$_4$N being deposited at 450 $^{\circ}$C. Equally important, micromagnetic simulations presented herein show that even with the presence of mixing layers, magnetic skyrmions, which arise from I-DMI, are found to be stable in MgO/Mn$_4$N/Pt$_x$Cu$_{1-x}$ thin films. This finding is consistent with recent experimental results. All these findings indicate that Mn$_4$N is a promising material for designing skyrmion-based memory.

\section{method}

We deposited 17 nm thick Mn$_4$N films on MgO substrates using a single Mn target by reactive radio frequency (rf) sputtering. The MgO substrates were wet-cleaned and heat-treated ex-situ \cite{Zhou2021_Mn4N}. The base pressure was 9 µPa. Sputter deposition was carried out at 450 $^{\circ}$C and under an Ar:N$_2$ gas flow ratio of 93:7. A Pt capping layer was deposited at room temperature to prevent oxidation. Our previous work showed epitaxially growth MgO/Mn$_4$N/Pt through X-ray diffraction (XRD) \cite{Zhou2021_Mn4N}

We performed X-ray photoelectron spectroscopy (XPS) measurements to obtain the compositional depth profile using the PHI VersaProbe III XPS instrument (Note: Certain commercial equipment is identified in this paper to foster understanding. Such identification does not imply recommendation or endorsement by NIST, nor does it imply that the materials or equipment identified are necessarily the best available for the purpose). XPS data were collected after sputtering off a few layers from the surface. Each sputtering lasted about 15 seconds, and the total sputtering time was 10 minutes. We analyzed the XPS results based on the following model with three assumptions. First, the composition from an XPS measurement is the averaged composition of a 3 nm layer from the surface, and the measurement sensitivity is the same within the 3 nm layer. Second, the distribution of Mn$_4$N near the Mn$_4$N/Pt and Mn$_4$N/MgO surfaces follows the cumulative distribution function (CDF) of a Gaussian  as below. 

\begin{equation} CDF(z; s, \sigma) = \frac{1}{\sigma\sqrt{2\pi}} \int_{-\infty}^{z} e^{-\frac{(t-s)^2}{2\sigma^2}} dt \end{equation} where z is the distance from the MgO substrate, s is the interface boundary and $\sigma$ is the thickness of the mixing.

Third, the distribution of the MnO is a Gaussian function (G) as below. 

\begin{equation} G(z ;w,l,\sigma) = \frac{w}{\sigma\sqrt{2\pi}}e^{-\frac{(z-l)^2}{2\sigma^2}} \end{equation} where w is the weight coefficient, l is the location of the MnO layers. 

\begin{table}[htb]
\caption{\label{tab:fittingXPS} The distribution functions of each layer used in XPS fitting, where CDF is the cumulative distribution function of a Gaussian distribution, G is the Gaussian function, and A, B, C, D, E are normalization coefficients.}

\begin{tabular}{l|l}
\hline
Layer & Distribution function \\
\hline
Pt & A(t) CDF(z; s$_1$ , $\sigma_1$)   \\
MnO at Mn$_4$N/Pt & B(t) G(z; w, l$_4$,  $\sigma_4$ )  \\
Mn$_4$N  & C(t) CDF(z; s$_2$ , $\sigma_2$)[1- CDF(z; s$_1$ , $\sigma_1$)] \\
MnO at MgO/Mn$_4$N & D(t) G(z; w, l$_5$,  $\sigma_5$ )  \\
MgO & E(t) CDF(z; s$_3$ , $\sigma_3$)  \\

\hline
\end{tabular}
\end{table}

Table 1 summarizes the distribution of each layer, where A, B, C, D, E are normalization coefficients. Using this model, we calculated the averaged compositions $f_{avr}(z)$ at location z by integrating the distribution from z – 3 nm to z and fitted these $f_{avr}(z)$ with the composition obtained from the XPS measurement by tuning parameters in CDF and G.

Polarized neutron reflectometry (PNR) measurements were performed at room temperature in a 3 T in-plane field using the PBR instrument at the NIST Center for Neutron Research. The reflectivities were measured as a function of the wavevector Q along the normal direction to the sample surface (i.e. specular scattering). Data were reduced and model-fitted with the reductus \cite{Maranville2018} and Refl1D \cite{KIRBY2012} software packages, respectively. The modeling allowed us to deduce the depth profiles of the film’s nuclear scattering length density ($\rho$$_N$, indicative of the nuclear composition), and the in-plane magnetization \cite{Majkrzak2006}. 

Micromagnetic simulations of Mn$_4$N were carried out using the Object-Oriented Micromagnetic Framework \cite{oommf}. Details of the simulations can be found in a previous publication \cite{Ma2021}. The simulation dimensions are 300 x 300 x 15 nm, and each cell is 5 x 5 x 5 nm. The exchange stiffness constant between each cell is 1.5 x 10-11 J/m. The external field is 0.01 T. The anisotropy (K$_u$) is (0.5 - 1.2) x 10$^5$ J/m$^3$. We added a 5 nm thick magnetically dead layer at the Mn$_4$N/Pt interface, which is assumed to provide a DMI according to the fractional composition of Pt in the layer but no magnetic interaction. Skyrmion size is defined as the average diameter of a boundary with zero magnetizations.

\section{Results and Discussion}

\begin{figure}
\label{fig:MH}
\includegraphics[width=0.5\textwidth]{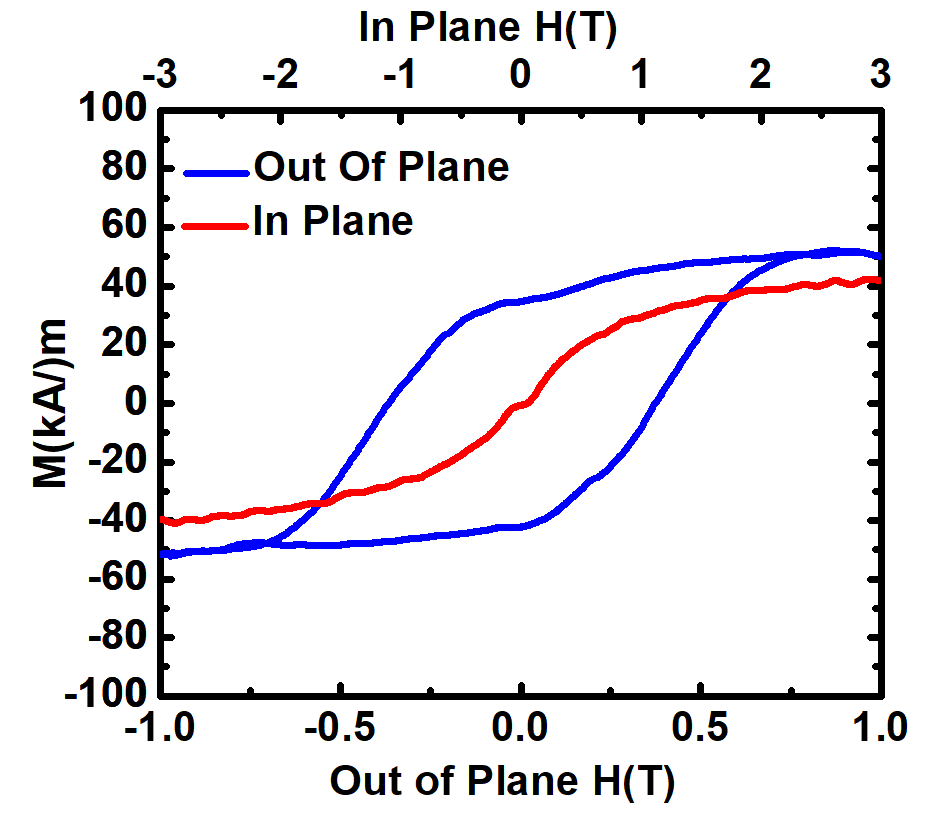}
\caption{Magnetic hysteresis loops of 17 nm thick Mn$_4$N at room temperature. The out-of-plane loop (blue) shows a coercivity of ~ 0.5 T and the in-plane loop (grey) shows an anisotropy field over 3 T. (1 T = 10000 Oe / $\mu_0$)}

\end{figure}

Figure 1 shows the hysteresis loops of Mn$_4$N at 300 K. The out-of-plane and in-plane M(H) loops confirm perpendicular magnetic anisotropy (PMA). The coercivity is found to be about 0.55 T, and saturation magnetization is 40 x kA/m. From the in-plane loop, the anisotropy field is determined to be over 3 T. The in-plane magnetization remains unsaturated at 3 T, which is the limit of the equipment. Using these results, the anisotropy (K$_u$) is estimated to exceed 70 kJ/m$^3$, comparable to other studies \cite{Yasutomi2014Mn,Kabara2015Mn4N,FOLEY2017Mnn4,Hirose2020Mn4N,HIROSE2020-1,Isogami2020Mn4n,Zhou2021_Mn4N}. Figure 2 shows (a) the compositional depth profile, and (b) the first derivative of the compositional depth profile, as a function of distance from the MgO substrate as determined from XPS. The surface of the MgO substrate is ideally set at z = 0 nm. In view of the presence of an interfacial layer, z = 0 nm is set at the position where the relative composition of Mg and O is 1:1. As such, z = 0 nm is the beginning of the MgO/Mn$_4$N interface region. The bulk of the Mn$_4$N layer is seen to extend to around 20 nm. Above approximately z = 20 nm is the Mn$_4$N/Pt interface region. From Figure 2 (a), elemental mixings exist at both MgO/Mn$_4$N and Mn$_4$N/Pt interface. At the MgO/Mn$_4$N interface region (z = 0 nm), both Mg and O atoms from the MgO substrate diffuse into Mn$_4$N. Since Mn reacts easily with O, but not Mg, O diffuses further into Mn$_4$N. Overall, the nominal average composition of this interface region is Mg$_{20}$O$_{37}$Mn$_{37}$N$_6$. From Figure 2 (b), the first derivative of the fitted compositional depth profile verifies that O diffuses further into the Mn$_4$N layer, as indicated by the peak of O occurring at higher z than that of Mg. Furthermore, the thickness of this interface region can be estimated using the full width at half maximum of the derivative. The mixing region with Mn$_4$N and MgO is about 5 nm thick, with O diffuses about 3 nm further into Mn$_4$N. The presence of this interfacial layer at the MgO/Mn$_4$N interface is possibly due to rough MgO/Mn$_4$N interfaces. From Shen et al. \cite{Shen2014}, misfit dislocations were observed by transmission electron microscopy, while such dislocations were not found at STO/Mn$_4$N interface \cite{Ghosh2021}.

\begin{figure}
\label{fig:XPS}
\includegraphics[width=0.5\textwidth]{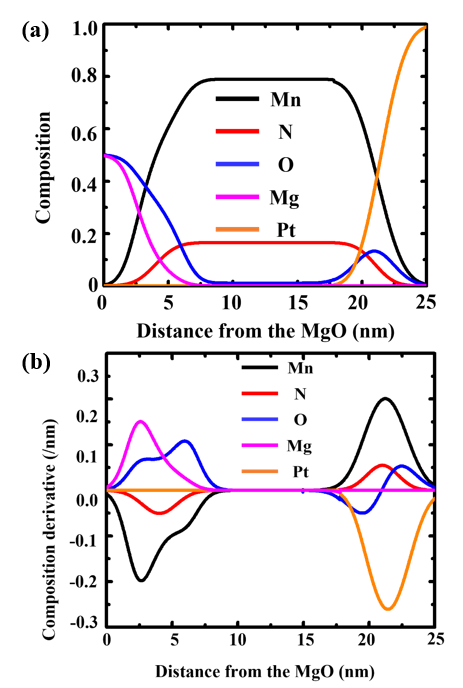}
\caption{(a) Compositional depth profile of MgO (001)/Mn$_4$N/Pt obtained from X-ray photoelectron spectroscopy (XPS) by fitting as a function of distance from MgO substrate. 0 nm is the surface of the MgO substrate. The fitted depth profile shows mixings at both MgO (001)/ Mn$_4$N and Mn$_4$N/Pt interface. (b) First derivative of the fitted depth profile of MgO (001)/Mn$_4$N/ as a function of distance from MgO substrate.}

\end{figure}

Both the Mn and N distributions shown in the composition derivative plot are separated by about 17 nm, which is the nominal thickness of the Mn$_4$N layer. It is worth noting that a small amount of O exists in the region. This is possibly due to the presence of O during the deposition. Even though the base pressure is below 9 µPa, and only Ar:N$_2$ was used, very tiny amounts of O can enter the system through the Ar or N$_2$ gas feed lines. Since Mn reacts easily with O, O enters Mn$_4$N. Overall, the composition of this region is Mn$_{77.7}$N$_{19.6}$O$_{2.7}$. This means that Mn:N maintains 4:1 in this sample, despite the presence of O. Above Mn$_4$N (z > 20 nm) lies the region of the Mn$_4$N/Pt interface. Both Pt and O atoms are found to diffuse into Mn$_4$N. The nominal average composition of this interface region is Mn$_{37}$N$_6$O$_{20}$Pt$_{37}$. The increase of O content at the Mn$_4$N/Pt interface is possibly due to the extended time required to cool down before room temperature deposition of Pt with trace amounts of O in the system. Using the full width at half maximum of Pt composition derivative in Figure 2 (b), this region is about 5 nm thick. It is worth noting that the mixing regions at both interfaces are about 5 nm thick, even though one was deposited at 450 $^{\circ}$C and Pt was deposited at room temperature.  

\begin{figure}
\label{fig:PNR}
\includegraphics[width=0.5\textwidth]{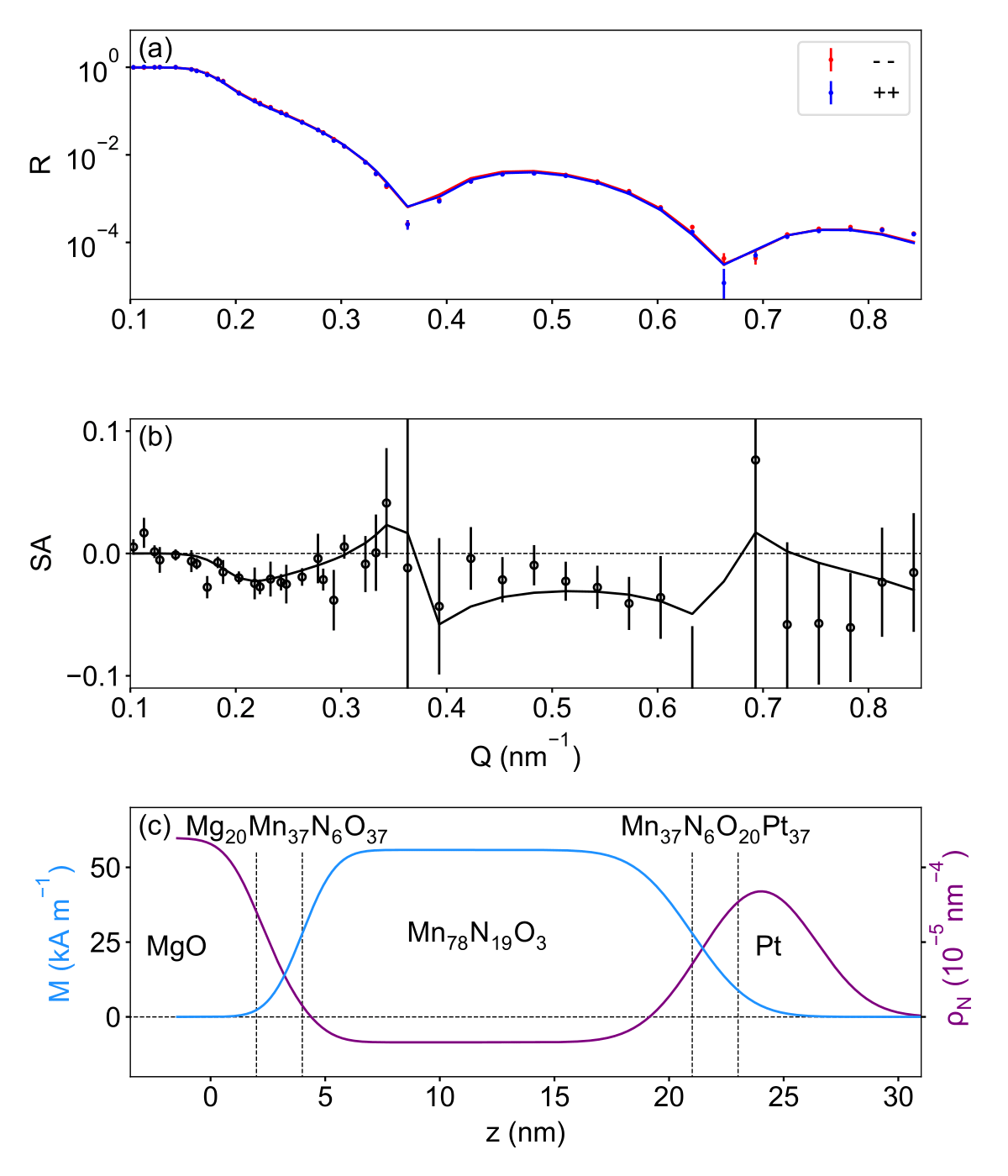}
\caption{(a) Model-fitted spin-dependent neutron reflectivities, R++ (blue) and R-- (red) measure at room temperature in 3 T. (b) Fitted data from (a) plotted as spin asymmetry to highlight the spin dependence. (c) In-plane magnetization (blue) and nuclear scattering length density (purple) depth profiles. Error bars in (a-b) correspond to 1 standard deviation.}

\end{figure}
To understand the magnetic consequences of the compositional inhomogeneity, we characterized the magnetization depth profile with PNR. Figure 3 (a) shows the model-fitted spin-dependent reflectivities (++ and --) as functions of wavevector transfer Q.  The data show clear oscillations, demonstrating sensitivity to the scattering length density depth profile.  However, the scattering is dominated by nuclear contributions, as the difference between ++ and -- is essentially too small to detect in (a).  In Fig. 3 (b) the same fitted data are plotted as spin-asymmetry (SA, difference in ++ and -- divided by the sum) which highlights magnetic contributions to the scattering.  Shown in this way, a spin-dependent signal is evident, demonstrating sensitivity to a non-zero magnetic depth profile.  Fig. 3(c) shows the magnetic and nuclear profiles corresponding to the best fit shown in panels (b) and (c).  In this model, the compositional profile of the nominal Mn$_4$N layer determined from XPS is approximated in terms of a 3 sublayer structure smeared out by error function roughness \cite{Maranville2018b} accounting for both conventional interlayer roughness and compositional gradients: MgO / Mg$_{20}$O$_{37}$Mn$_{37}$N$_6$ / Mn$_{78}$N$_{19}$O$_3$ / Mn$_{37}$N$_6$O$_{20}$Pt$_{37}$ / Pt.  All nuclear scattering length densities were calculated based on the expected composition and were treated as fixed parameters in the fitting.  Mg$_{20}$O$_{37}$Mn$_{37}$N$_6$ and Mn$_{37}$N$_6$O$_{20}$Pt$_{37}$ sublayer thicknesses were held fixed at 2 nm, and the Pt cap layer thickness was fixed at 3 nm.  Interlayer roughness was assumed to be equal for all interfaces, and it was found to be 1 nm. The Mn$_{78}$N$_{19}$O$_{3}$ layer thickness was a free parameter, found to be 17 nm, confirming XPS results.  The magnetizations of all 3 sublayers of the nominal Mn$_4$N layer were treated as free parameters. The profile shows a center Mn$_{78}$N$_{19}$O$_{3}$ layer magnetization of 56 kA / m, with a sharp decline near both interfaces.  While the magnetization does not go completely to zero in the interfacial sublayer regions, this is consistent with layers of zero magnetization smeared out by the apparent roughness \cite{Kirby2018}.  The integrated magnetization for the entirety of the nominal Mn$_4$N film is 45 kA/m, consistent with the magnetometry data shown in Fig. 1.  Despite the rigid constraints, Fig. 3 (a-b) show that this model gives an excellent fit to the data, and we note that models with a uniform magnetization profile in the nominal Mn$_4$N layer result in a significantly worse fit.  Thus, the PNR results confirm the compositional profile determined from XPS, and show that the interfaces of the nominal Mn$_4$N layers contribute no net magnetization.  This finding is crucial in designing spintronics devices with Mn$_4$N. With non-magnetic interfacial mixings, these regions can be treated as a non-magnetic layer for the effectiveness of spin currents and spin-orbit torque in Mn$_4$N heterostructures. Thus, Mn$_4$N remains a viable option for future devices. 

\begin{figure}
\label{fig:sims}
\includegraphics[width=0.5\textwidth]{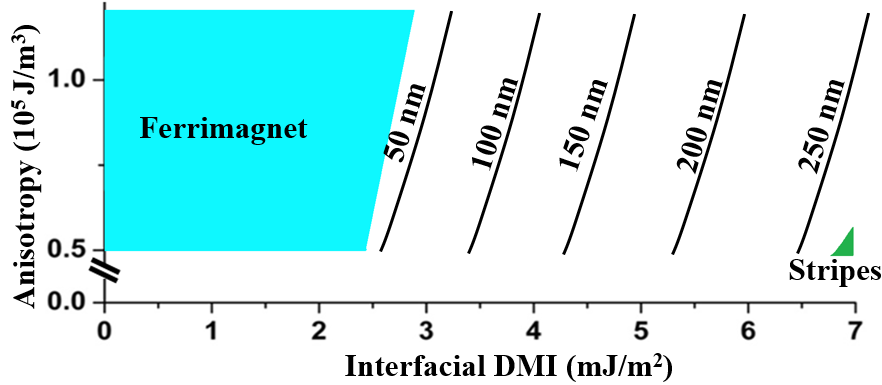}
\caption{Micromagnetic simulation of skyrmions in 15 nm Mn$_4$N with 5 nm of magnetic dead layer separated the interface and the Mn$_4$N sample at 300 K. Anisotropy (K$_u$) varies from 0.5 to 1.2 x 10$^5$ J/m$^3$. Interfacial DMI varies from 0 to 7 mJ/m$^2$. The blue region corresponds to ferrimagnetic states. Black lines correspond to the size of skyrmions in Mn$_4$N. The green region corresponds to larger than 250 nm skyrmions which form into stripes because of the limited simulated space of 300 nm x 300 nm used.}

\end{figure}

Micromagnetic simulations are employed to study the effect of interfacial mixings on skyrmion formation in Mn$_4$N. Figure 4 shows the DMI-K$_u$ phase diagram of a 15 nm thick Mn$_4$N with a 5 nm thick magnetic dead layer at 300 K. Only the Mn$_4$N/Pt mixing region is considered here. In the MgO/Mn$_4$N/Pt heterostructure, interfacial-DMI is dominated by Pt, so a magnetic dead layer at Mn$_4$N/Pt can potentially influence skyrmion formation. From XPS, this region has an average composition of  Mn$_{37}$N$_6$O$_{20}$Pt$_{37}$. From density functional theory (DFT) calculations, interfacial-DMI at Mn$_4$N/Pt is near |7| mJ/m$^2$. Using a simple fractional composition model, this region is estimated to have a DMI of 2.6 mJ/m$^2$. On the other hand, little or no effect is expected for the MgO/Mn$_4$N as the interfacial-DMI is negligible between MgO and Mn \cite{Ma2021}. Thus, this region is treated as a non-interacting empty space. Using experimental K$_u$ of 0.7 x 10$^5$ J/m$^3$, robust skyrmions are observed in interfacial-DMI above 2.5 mJ/m$^2$. Below this DMI, only ferrimagnetic states are observed. With interfacial-DMI of 3.5 mJ/m$^2$, near 100 nm skyrmions are found in Mn$_4$N. As interfacial-DMI increases to 5.5 mJ/m$^2$, skyrmion sizes increase to 200 nm. At interfacial-DMI close to 7.0 mJ/m$^2$, 250 nm skyrmions are observed in the simulations. With larger interfacial-DMI, skyrmions start to evolve into stripes. However, this phenomenon could be due to the limit of the simulation space of 300 nm in the in-plane direction.  

From Figure 4, using experimental K$_u$ of 0.7 mJ/m$^3$ and calculated interfacial-DMI at Mn$_4$N/Pt of |7| mJ/m$^2$, skyrmions are just over 250 nm in size. This is comparable to the observed skyrmions in MgO/Mn$_4$N/Pt, which range from 250 nm to 350 nm in size. Furthermore, from experiment, skyrmions in MgO/Mn$_4$N/Pt$_{0.1}$Cu$_{0.9}$ are around 100 nm. From DFT calculation, the interfacial-DMI of Cu/Mn$_4$N is 2.6 mJ/m$^2$ \cite{Ma2021}. Assuming a simple solution model, the interfacial-DMI of Mn$_4$N/Pt$_{0.1}$Cu$_{0.9}$ is estimated to be 3.0 mJ/m$^2$. From Figure 4, with K$_u$ of 0.7 mJ/m$^3$ and interfacial-DMI of 3.0 mJ/m$^2$, skyrmions are about 60 nm in diameter. This is smaller than the measured skyrmions in MgO/Mn$_4$N/Pt$_{0.1}$Cu$_{0.9}$, which are 80 to 120 nm in size. This discrepancy may be due to the crude model to estimate the DMI in a mixture. Figure 5 shows the color mapping of out-of-plane magnetization at various K$_u$ (0.5, 0.7, and 0.9 mJ/m$^3$) and interfacial-DMI (1.5, 3.0, 4.5, and 6.0 mJ/m$^2$), with blue regions corresponding to up magnetic moments and red regions corresponding to down magnetic moments. Robust shapes of skyrmions are found in this Mn$_4$N heterostructures throughout various K$_u$ and interfacial-DMI, even with the presence of interfacial mixings. This is critical for any potential future for using Mn$_4$N in spintronics applications. Overall, these results verify the reliability of Mn$_4$N to host skyrmions, despite the existence of magnetic dead layers, and confirm the potential of using Mn$_4$N in skyrmion-based devices.

\begin{figure}
\label{fig:simsmap}
\includegraphics[width=0.5\textwidth]{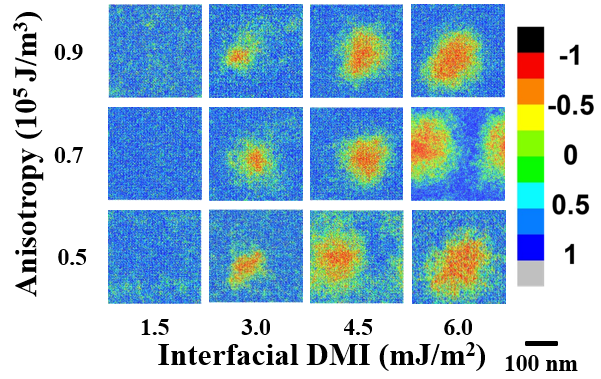}
\caption{Color mapping of out-of-plane magnetization in 15 nm Mn$_4$N with magnetically dead layers at 300 K from micromagnetic simulation. Red regions correspond to down magnetic moments and blue regions correspond to up magnetic moments.}
\end{figure}

\section{Conclusion}

Ferrimagnetic Mn$_4$N thin films are recently recognized as promising skyrmionic materials due to thermal stability and the ability to host small skymions. These material attributes prompted us to study interfacial mixing effect in MgO/Mn$_4$N/(Pt-Cu) heterostructures. From X-ray photoelectron spectroscopy, elemental mixings were revealed at both MgO/Mn$_4$N and Mn$_4$N/Pt interface. These mixing regions are about 5 nm thick, comparable to room temperature deposition. Using polarized neutron reflectometry, we found that these interface regions exhibit essentially zero magnetization. To address the effect of these inhomogeneities on the potential of using Mn$_4$N in spintronic devices, micromagnetic simulations were employed to investigate skyrmions in Mn$_4$N. Simulated skyrmions were found to be robust at room temperature, despite the existence of magnetic dead layers. These results confirm that Mn$_4$N is a promising material for next-generation memory and logic devices. 

\section*{Acknowledgments}

This work was partially supported by the DARPA Topological Excitations in Electronics (TEE) program (grant D18AP00009). The content of the information does not necessarily reflect the position or the policy of the Government, and no official endorsement should be inferred. Approved for public release; distribution is unlimited. The Phi VersaProbe III XPS used for acquiring the data was provided through the NSF-MRI Award \#1626201.

\section*{DATA AVAILABILITY STATEMENT}

The data that support the findings of this study are available from the corresponding author upon reasonable request.

\bibliography{ref}

\end{document}